\documentclass[conference]{llncs}
\usepackage{amsmath,amssymb,amsfonts}
\usepackage{graphicx}
\usepackage{textcomp}
\usepackage{xcolor}
\usepackage[nobiblatex]{xurl}
\usepackage{hyperref}
\usepackage{empheq}
\usepackage{algpseudocode}
\def\BibTeX{{\rm B\kern-.05em{\sc i\kern-.025em b}\kern-.08em
T\kern-.1667em\lower.7ex\hbox{E}\kern-.125emX}}

\usepackage{standalone}

\usepackage{hyperref}
\usepackage{cleveref}
\crefname{algocf}{alg.}{algs.}
\Crefname{algocf}{Algorithm}{Algorithms}
\crefname{AlgoLine}{lin.}{lins.}
\Crefname{AlgoLine}{Line}{Lines}

\usepackage{url}

\usepackage{subcaption}

\usepackage{enumitem}
\usepackage[ruled,linesnumbered,commentsnumbered,noend]{algorithm2e}
\usepackage{tabulary}

\newcommand{\universe}[0]{\mathcal{U}}
\newcommand{\actuniverse}[0]{\mathcal{A}} 
\newcommand{\oceluniverse}[0]{\universe_{\eventlogvar}}

\newcommand{\otuniverse}[0]{\universe_{\mathit{ot}}}
\newcommand{\oiuniverse}[0]{\universe_{\mathit{oi}}}

\newcommand{\eventuniverse}[0]{\universe_{ev}}

\newcommand{\set}[1]{\{ #1 \}} 
\newcommand{\multiset}[1]{[\hspace{0.1em} #1 \hspace{0.1em}]} 
\newcommand{\seq}[1]{\langle #1 \rangle} 

\newcommand{\powerset}[1]{\mathcal{P}(#1)}
\newcommand{\allsequence}[1]{#1^*}
\newcommand{\proj}[2]{#1_{\upharpoonright_{#2}}} 
\newcommand{\partmapsto}[0]{\not\to} 
\newcommand{\st}[0]{\mid} 
\newcommand{\length}[1]{\vert #1 \vert} 
\newcommand{\seqvar}[0]{\sigma}

\newcommand{\gettuple}[2]{#1.#2}

\newcommand{\eventvar}{e}

\newcommand{\eidvar}{\mathit{ei}}
\newcommand{\eactvar}{\mathit{act}}
\newcommand{\etimevar}{\mathit{time}}
\newcommand{\eomapvar}{\mathit{omap}}
\newcommand{\evmapvar}{\mathit{vmap}}

\newcommand{\tracevar}[0]{\rho}
\newcommand{\caseidvar}[0]{\mathit{cid}}
\newcommand{\eventlogvar}[0]{L}
\newcommand{\eventlogeventsvar}{E}
\newcommand{\eventorder}{\preceq_{\eventlogeventsvar}}

\newcommand{\eventsetvar}[0]{E}

\newcommand{\eventmapping}{\pi}

\newcommand{\geteventattr}[2]{\eventmapping_{#2}(#1)}

\newcommand{\otvar}{\mathit{ot}}
\newcommand{\otsetvar}{\mathit{OT}}
\newcommand{\oteventlog}[1]{\otsetvar_{#1}}

\newcommand{\enhancedplacenetsset}[0]{\mathit{pt}}

\newcommand{\placevar}[0]{p}
\newcommand{\placenetvar}[0]{\netvar_{\placevar}}

\newcommand{\lpmsetvar}[0]{\mathit{LPM}}
\newcommand{\lpmvar}[0]{\mathit{lpm}}

\newcommand{\oclpmvar}{\mathit{oclpm}}
\newcommand{\oclpmsetvar}{\mathit{OCLPM}}
\newcommand{\netvar}{N}
\newcommand{\ocpnvar}{\mathit{ON}}
\newcommand{\netarcsvar}{F}
\newcommand{\netvararcsvar}{F_{\mathit{var}}}
\newcommand{\placetypevar}{\mathit{pt}}

\newcommand{\placeoraclefunc}{\mathit{po}}
\newcommand{\processexecutionfunc}[0]{\mathit{peo}}
\newcommand{\flatocelfunc}[0]{\mathit{flat}}

\newcommand{\vararcscorefunc}[0]{\mathit{score}}
\newcommand{\variablearcfunc}[0]{\mathit{vararc}}

\newcommand{\constname}[1]{\textit{#1}}

\newcommand{\lpmtext}[0]{LPM}
\newcommand{\oclpmtext}[0]{OCLPM}
\newcommand{\oceltext}[0]{OCEL}

\newcommand{\ocpntext}{OCPN}

\newcommand{\undefined}[0]{\perp}

\begin{document}

    \title{Object-Centric Local Process Models
    \thanks{We thank the Alexander von Humboldt (AvH) Stiftung for supporting our research.}
    }

    \author{Viki Peeva\and
    Marvin Porsil \and
    Wil M.P. van der Aalst}
    \authorrunning{V. Peeva et al.}
%
    \institute{Chair of Process and Data Science \\ RWTH Aachen University, Aachen, Germany \\
    \email{\{peeva,wvdaalst\}@pads.rwth-aachen.de} \\
    \email{marvin.porsil@rwth-aachen.de}}

    \maketitle

    \begin{abstract}
        Process mining is a technology that helps understand, analyze, and improve processes.
        It has been present for around two decades, and although initially tailored for business processes, the
        spectrum of analyzed processes nowadays is evermore growing.
        To support more complex and diverse processes, subdisciplines such as object-centric process mining and
        behavioral pattern mining have emerged.
        Behavioral patterns allow for analyzing parts of the process in isolation, while object-centric process
        mining enables combining different perspectives of the process.
        In this work, we introduce \emph{Object-Centric Local Process Models} (OCLPMs).
        OCLPMs are behavioral patterns tailored to analyzing complex processes where no single case notion exists and
        we leverage object-centric Petri nets to model them.
        Additionally, we present a discovery algorithm that starts from object-centric event logs, and implement the
        proposed approach in the open-source framework ProM.
        Finally, we demonstrate the applicability of OCLPMs in two case studies and evaluate the approach on various
        event logs.
    \end{abstract}


    \section{Introduction}\label{sec:introduction}
    Process mining takes event data generated as a byproduct of organizations' operations and provides insights and
    improvements of the analyzed process.
    This is achieved by automatically discovering process models, computing conformance checking
    metrics, or enhancing the model with concrete KPIs.
    Traditional process mining considers the process from start to end and uses a single case notion.
    However, in reality, the process interacts with various entities, in the community known as object types or
    artifacts.
    There exist different strategies how to connect or model such entities together with the control-flow of the process.
    In our work, we focus on \emph{object-centric process mining} as described in~\cite{DBLP:journals/fuin/AalstB20}
    and \emph{Object-Centric Event Logs} (\oceltext{}s)~\cite{DBLP:conf/adbis/GhahfarokhiPBA21}.
    Moreover, process issues like delays, high costs, etc., almost never occur on a global level but in specific
    subcontexts, requiring pattern mining.
    Pattern mining is a known discipline in data science and the concept has also been established in the area of process
    mining with works for discovering frequent subsequences~\cite{DBLP:conf/icde/AgrawalS95},
    episodes~\cite{DBLP:conf/simpda/LeemansA14a}, and local process models~\cite{DBLP:journals/jides/TaxSHA16}.
    While end-to-end process models describe the entire process, process or behavioral patterns only explain (match)
    particular sub-behaviors of the process.
    In particular, the proposed approach focuses on \emph{Local Process Models} (\lpmtext{}s), which are a type of
    behavioral pattern, allowing for constructs such as sequence, choice, concurrency, and loop.

    To combine the two areas, in this paper, we define \emph{Object-Centric Local Process Models} (\oclpmtext{}s) as a
    behavioral pattern alternative for object-centric process mining.
    Additionally, we build a framework around an already existing \lpmtext{} discovery approach to discover such \oclpmtext{}s.
    The discovery approach is built upon the formalisms of Petri nets, and as a result, the discovered \oclpmtext{}s are
    represented as object-centric Petri nets (\ocpntext{}s).
    Specifically, we list the following contributions:
    \begin{itemize}
        \item [(1)] Adapting existing \lpmtext{} discovery approach for \oclpmtext{} discovery.
        \item [(2)] Implementing the algorithm in the publicly accessible framework ProM.
        \item [(3)] Demonstrating feasibility and applicability in real-world scenarios.
    \end{itemize}

    The rest of the paper is structured as follows.
    First, we illustrate the necessity for \oclpmtext{}s in \Cref{sec:motivating-example}.
    Then, we present related work in \Cref{sec:related_work}, and give the necessary background to follow the rest of
    the paper in \Cref{sec:preliminaries}.
    In \Cref{sec:method}, we describe the proposed framework and all the surrounding details, after which,
    \Cref{sec:evaluation} covers the experiments demonstrating its applicability.
    In \Cref{sec:discussion}, we discuss the strengths and weaknesses of the proposed approach.
    Finally, in \Cref{sec:conclusion} we conclude the paper and offer an outlook on possible extensions.

    \section{Motivating Example}\label{sec:motivating-example}
\begin{figure}[t]
        \centering
        \includegraphics[width=\linewidth]{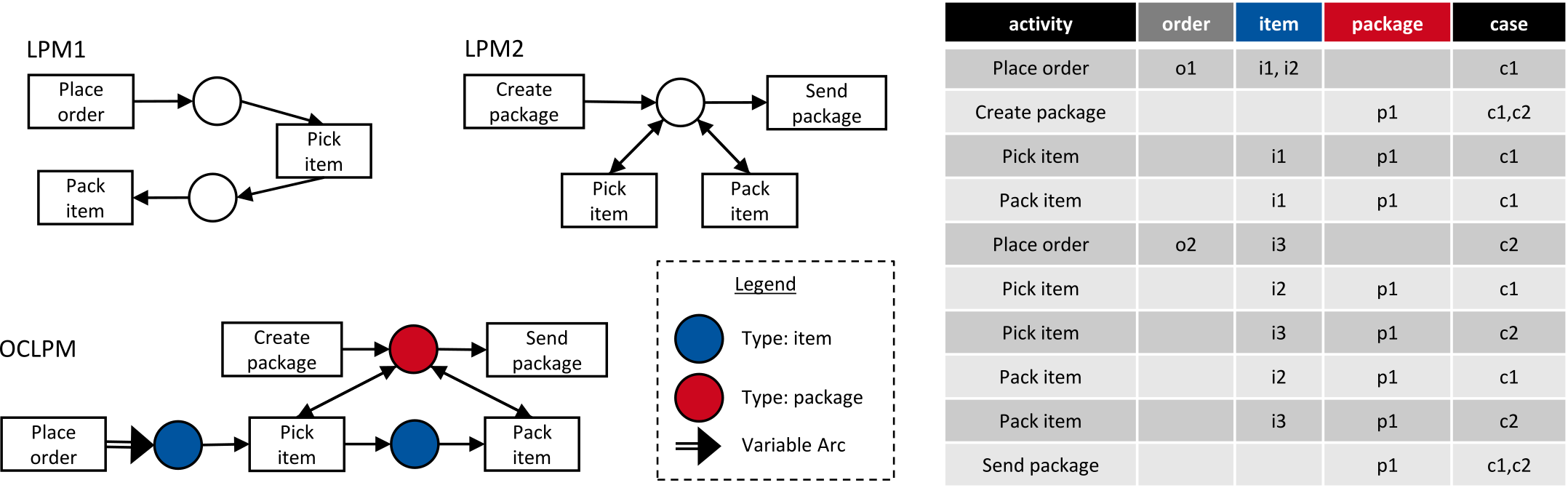}
        \caption{Event log exceprt with example \lpmtext{}s from the perspectives of the items ($\lpmvar1$) and packages
            ($\lpmvar2$) and one \oclpmtext{} depicting both perspectives.}
        \label{fig:motivational_example}
    \end{figure}

    The necessity of pattern mining for processes has been demonstrated and discussed in previous
    works~\cite{DBLP:journals/jides/TaxSHA16,DBLP:conf/apn/PeevaMA22}.
    Challenges like spaghetti and flower process models make behavioral patterns even more attractive.
    However, current pattern representations lack the ability to model the process from multiple viewpoints.
    We use the example depicted in \Cref{fig:motivational_example} to show the benefits of having patterns that are
    object-centric aware.
    The excerpt event log is for an order management process and includes ten events of five different activities and three
    object types.
    To discover traditional \lpmtext{}s on such object-centric event logs, we would choose one object type and focus on
    the viewpoint of the chosen object type.
    From the perspective of each item, the process starts with \emph{Place order}.
    However, in the process, one \emph{Place order} is executed for more items, meaning one \emph{Place order} event is
    followed by multiple \emph{Pick item} and \emph{Pack item} events.
    This can not be caught by the model because of replicating events, also called \emph{convergence}.
    Moreover, for each item, first \emph{Pick Item} and then \emph{Pack Item} occurs.
    However, from the perspective of the package, it would appear there are random interleaving of the two activities.
    Therefore, resulting in unconnected loops in the model, see $\lpmvar2$ in \Cref{fig:motivational_example}, called
    \emph{divergence}.

    By using \oclpmtext{}s and modeling the multiple perspectives together resolves the aforementioned problems.
    Consider the \oclpmtext{} in \Cref{fig:motivational_example}.
    The execution of \emph{Place order} resulting in multiple items is clearly depicted with the variable arc between
    \emph{Place order} and \emph{Pick item}, solving the convergence problem.
    Additionally, the order of \emph{Pick item} and \emph{Pack item} is explicitly represented in the \oclpmtext{}, avoiding
    divergence.
    Considering this, we conclude that \oclpmtext{}s are valuable for processes with multiple object types.

    \section{Related Work}\label{sec:related_work}
    \textbf{Object-centric modeling.} As indicated in the introduction, there exist multiple strategies for modeling
    control-flow together with the associated data participators.
    Artifact-centric approaches~\cite{DBLP:conf/bpm/BhattacharyaGHLS07,DBLP:journals/debu/CohnH09,DBLP:journals/tsc/LuNWF15} model the business process in terms of business artifacts.
    Such artifacts contain data and change states.
    A state change can result because of a step in the process, or it can trigger an advance in the process.
    Flexible case modeling~\cite{DBLP:conf/bpm/HeweltW16,DBLP:conf/caise/HaarmannMW21} breaks the process into a domain
    model defining the participating entities, object lifecycles, process fragments to model the control-flow, and a
    goal state.
    Proclets~\cite{DBLP:conf/coopis/AalstBEW00,DBLP:conf/apn/Fahland19} model the control flow from the perspective of
    the different object types and can have synchronization points between them.
    The synchronization points allow for one-to-one and one-to-many relationships.
    Although powerful, not many techniques in process mining are able to work with such models.
    Event-knowledge graphs and object-centric process models~\cite{DBLP:books/sp/22/Fahland22,DBLP:journals/fuin/AalstB20} are newer paradigms that put the control-flow in the main focus, but now from the
    perspective of multiple key entities, known as object types.
    For more information regarding the different modeling strategies, we refer to~\cite{DBLP:conf/apn/Fahland19,DBLP:journals/fuin/AalstB20}.
    However, the more involved modeling strategies can be quite complex, and the ones focusing on control-flow can
    result in spaghetti models.

    \textbf{Behavioral patterns.} All modeling strategies above, represent the entire process in one model.
    However, processes can be complex, resulting in complicated over-fitting, spaghetti, models or simple, but
    under-fitting, flower, models.
    Some approaches, alleviate the complexity by introducing scenarios or fragmenting the
    process~\cite{DBLP:conf/apn/Fahland09,DBLP:conf/IEEEscc/HaarmannLW22}.
    However, these still have the goal of representing the entire process.
    Behavioral patterns, like sequences, episodes, and local process
    models~\cite{DBLP:conf/icde/AgrawalS95,DBLP:journals/datamine/MannilaTV97,DBLP:journals/jides/TaxSHA16}, deal with
    complexity by modeling subparts of the process and ignoring the rest.
    This makes them suitable for a spectrum of applications and not only modeling the process
    (see~\cite{DBLP:conf/apn/PeevaMA22}).
    An alternative are declarative process modeling languages~\cite{DBLP:journals/ife/AalstPS09} that instead of using
    the activities to model control-flow, they define a set of constraints between the activities, like precedence or
    response.
    There also exist object-centric extensions for declarative constraints~\cite{DBLP:conf/bis/LiCA17}, where constraints
    are related to object types.
    However, with this work, we extend local process models as in~\cite{DBLP:conf/apn/PeevaMA22} to be
    object-centric~\cite{DBLP:journals/fuin/AalstB20}.

    \section{Preliminaries}\label{sec:preliminaries}
    We define sets ($X = \set{a,b}$), multisets ($ M = [a^2,b^3]$), sequences ($\seqvar = \seq{a,b,c}$ where $\seqvar_1
    = a$), and tuples ($t = ( a,b,c )$).
    Given a set $X$, $\powerset{X}$ is the power set of $X$, and $\allsequence{X}$ represents the set of all sequences
    over $X$.
    We use $f(X)=\{f(x) \st x \in X\}$ ($f(\seqvar)=\langle f(\seqvar(1)), f(\seqvar(2)),\dots, f(\seqvar(n) \rangle$)
    to apply the function $f$ to every element in the set $X$ (the sequence $\seqvar$).
    Finally, we write $\proj{f}{X}$ ($\proj{\seqvar}{X}$) to denote the projection of the domain of function $f$ (the
    sequence $\seqvar$) onto $X$.

    \subsubsection{Event Logs}\label{subsec:preliminiaries-logs}
    Collected data used for process analysis is transformed into \emph{event logs}.
    In \Cref{def:event}, we define events, and in \Cref{def:event-log}, we formally define \emph{event logs} that can
    be used for both traditional and object-centric processes.

    \begin{definition}[Event] \label{def:event}
    Let $\eventuniverse$ be the universe of events, $\otuniverse$ the universe of object types, and $\oiuniverse$ the
    universe of object identifiers.
    We define the event $\eventvar = (\eidvar, \eactvar, \etimevar, \eomapvar, \evmapvar)$, such that $\geteventattr{
        \eventvar}{\eidvar} = \eidvar$ is the event id, $\geteventattr{\eventvar}{\eactvar} = \eactvar$ is the event
    activity, $\geteventattr{\eventvar}{\etimevar} = \etimevar$ is the timestamp of the event,
    $\geteventattr{\eventvar}{\eomapvar} = \eomapvar$ ($\eomapvar \in \otuniverse \to \powerset{\oiuniverse}$) is a
    function mapping each object type to the objects involved in the event,
    and $\geteventattr{\eventvar}{\evmapvar} = \evmapvar$ is a function assigning values to each of the event attributes.
    \end{definition}

    \begin{definition}[Event Log~\cite{DBLP:journals/fuin/AalstB20}] \label{def:event-log}
    $\eventlogvar = (\eventlogeventsvar, \eventorder) \in \oceluniverse$ is an event log with $\eventlogeventsvar
    \subseteq \eventuniverse$ and $\eventorder \subseteq \eventlogeventsvar \times \eventlogeventsvar$
    such that:
    \begin{itemize}
        \item $\eventorder$ defines a partial order (reflexive, antisymmetric, and transitive)
        \item $\forall_{\eventvar_1, \eventvar_2 \in \eventlogeventsvar} \, \geteventattr{\eventvar_1}{\eidvar} =
        \geteventattr{\eventvar_2}{\eidvar} \implies \eventvar_1 = \eventvar_2$, and
        \item $\forall_{\eventvar_1, \eventvar_2 \in \eventlogeventsvar} \, \eventvar_1 \eventorder \eventvar_2 \implies
        \geteventattr{\eventvar_1}{\etimevar} \leq \geteventattr{\eventvar_2}{\etimevar}$.
    \end{itemize}
    We use $\oteventlog{\eventlogvar}$ to denote all object types in \eventlogvar.
    \end{definition}

    In process mining, the concept of process executions is essential for many techniques.
    Depending on the complexity, process executions are represented as totally or partially ordered events.
    For example, in traditional process mining, event logs are represented as a set of traces, where each trace is a
    sequence of events.
    Here, we call such event logs \emph{simple event logs} and write $\eventlogvar_S \in \powerset{\allsequence{
        \eventuniverse}}$.
    To obtain a simple event log $\eventlogvar_S$ from $\eventlogvar = (\eventlogeventsvar, \eventorder) \in \oceluniverse$,
    we group the events on a certain object type $ot \in \otuniverse$ and order them.
    For this, we use $\flatocelfunc \in \oceluniverse \times \otuniverse \to \powerset{\allsequence{\eventuniverse}}$
    such that $\flatocelfunc(\eventlogvar, \otvar) = \set{\tracevar_{\caseidvar} \in \allsequence{\eventuniverse} \st
    \caseidvar \in \bigcup_{\eventvar \in \eventlogeventsvar} \geteventattr{\eventvar}{\eomapvar}(\otvar) \land
    \forall_{\eventvar \in \eventlogeventsvar}
        (\caseidvar \in \geteventattr{\eventvar}{\eomapvar}(\otvar) \implies \eventvar \in \tracevar) \land
        \forall_{1 \leq i < j \leq \length{\tracevar}}
        (\geteventattr{\tracevar_i}{\etimevar} \leq \geteventattr{\tracevar_j}{\etimevar})}$.
    In case two events have the same timestamp, we assume some order.

    \subsubsection{Process Models}\label{subsec:preliminiaries-models}
    The behavior recorded in logs can be modeled using different notations, such as DFG, process trees, BPMN, Petri
    nets, etc.
    In this work, we focus on Petri nets, and more precisely on \emph{labeled Petri nets} which we define in
    \Cref{def:labeled-petri-net} and \emph{object-centric Petri nets} as defined in \Cref{def:ocpn}.

    \begin{definition}[Labeled Petri Net]
        \label{def:labeled-petri-net}
        A \emph{labeled Petri net} is a tuple $N = (P,T,F,l)$ with $P$ the set of places, $T$ the set of transitions, such
        that $P \cap T = \emptyset$, $F \subseteq (P \times T) \cup (T \times P)$ the flow relation and
        $l \in T \to \actuniverse \cup \set{\tau}$ a labeling function.
    \end{definition}

    \begin{definition}[Object-centric Petri nets]
        \label{def:ocpn}
        An \emph{Object-centric Petri net} is a tuple $\ocpnvar = (\netvar, \placetypevar, \netvararcsvar)$
        where $\netvar = (P, T, F, l)$ is a labeled Petri net, $\placetypevar \in P \to \otuniverse$ maps places onto
        object types, and $\netvararcsvar \subseteq \gettuple{\netvar}{\netarcsvar}$ is the subset of variable arcs.
    \end{definition}

    Generally, process models define a language, commonly used in process mining to measure how well the
    model aligns with the collected event data.
    For both labeled and object-centric Petri nets, exist notions like tokens and markings, necessary to define their language.
    Because of space restrictions, we refer to~\cite{DBLP:journals/fuin/AalstB20} on how the language is
    obtained.

    In this work, we consider \lpmtext{}s as labeled Petri nets and \oclpmtext{}s to be \ocpntext{}s.
    However, \lpmtext{}s and \oclpmtext{}s do not cover the entire event log, since they do not represent the entire
    process.
    Therefore, we define $\eventsetvar_{\oclpmvar} \subseteq \eventlogeventsvar$ to be the events in the event log
    $\eventlogvar = (\eventlogeventsvar, \eventorder)$ that are covered by the \oclpmtext{} $\oclpmvar$.
    More details about how \lpmtext{}s are matched to event logs, can be found in previous
    works~\cite{DBLP:journals/jides/TaxSHA16,DBLP:conf/apn/PeevaMA22}.

    \section{OCLPM Discovery}\label{sec:method}
\begin{figure}[t]
        \centering
        \includegraphics[width=\linewidth]{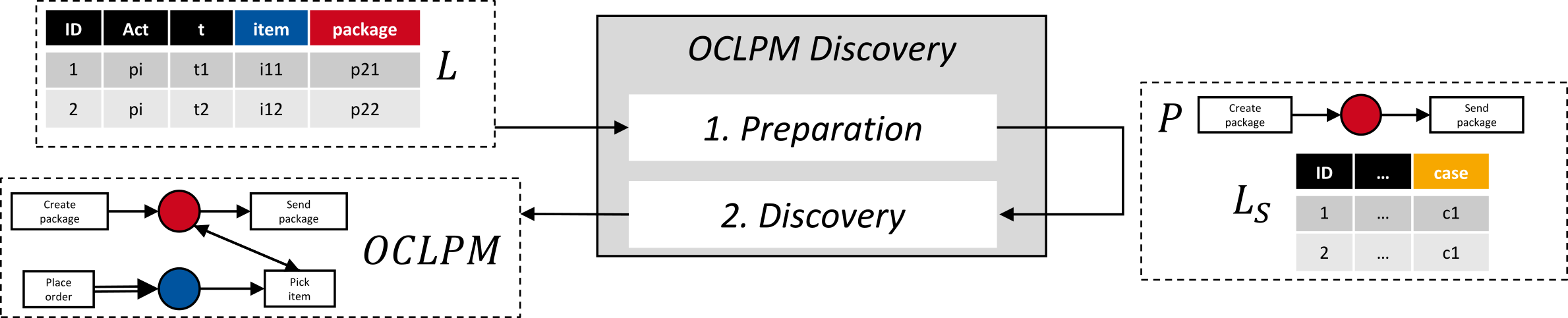}
        \caption{Overview of the steps in the \oclpmtext{} framework, depicted with their inputs and outputs.}
        \label{fig:framework}
    \end{figure}

    In this section, we present a two-phase discovery approach for \oclpmtext{}s given an \oceltext{}.
    In \Cref{fig:framework} we visualize the two phases of the approach: \emph{preparation} and \emph{discovery},
    together with the input, output, and the intermediate results.
    In \Cref{alg:discovery} we sketch the steps of the algorithm.

    \begin{algorithm}
        \SetKwFunction{PlaceOracle}{po}\SetKwFunction{Flatten}{flat}\SetKwFunction{ProcessExecutionOracle}{peo}
        \SetKwFunction{LPMDiscovery}{lpmd}\SetKwFunction{VarArcIdentification}{vararc}
        \SetKwData{Log}{$\eventlogvar$}\SetKwData{SimpleLog}{$\eventlogvar_S$}
        \SetKwData{Places}{$P$}\SetKwData{PlaceTypes}{$PT$}
        \SetKwData{Lpms}{$\lpmsetvar$}\SetKwData{Oclpms}{$\oclpmsetvar$}
        \SetKwInOut{Input}{input}\SetKwInOut{Output}{output}

        \Input{$\Log = (\eventsetvar, \eventorder)$}

        \tcp{Preparation (Phase 1)}
        $\PlaceTypes \leftarrow \bigcup_{\otvar \in \oteventlog{\Log}}\PlaceOracle(\Flatten(\Log, \otvar)) \times \set{ot
        }$\;\label{algline:discover_place_types}
        $\SimpleLog \leftarrow \Flatten(\ProcessExecutionOracle(\Log))$\;\label{line:flatten_log}

        \tcp{Discovery (Phase 2)}
        $\Places \leftarrow \set{\placenetvar \st (\placenetvar, \otvar) \in \PlaceTypes}$\;
        $\Lpms \leftarrow \LPMDiscovery(\SimpleLog, \Places)$\;
        $\Oclpms \leftarrow \emptyset$\;
        \For{$\lpmvar \in \Lpms$}{
            $\placetypevar \leftarrow \set{(p, \otvar)
                \st p \in \gettuple{\lpmvar}{P} \land (\proj{\lpmvar}{p}, \otvar) \in \PlaceTypes}$\footnotemark\;
            $\netvararcsvar \leftarrow \VarArcIdentification(\lpmvar, \placetypevar, \SimpleLog)$\;
            $\Oclpms \leftarrow \Oclpms \cup \set{(\lpmvar, \placetypevar, \netvararcsvar)}$\;
        }
        \KwRet{\Oclpms}
        \caption{Discovery algorithm for \oclpmtext{}s.}\label{alg:discovery}
    \end{algorithm}

    \subsection{Preparation (Phase 1)}\label{subsec:preparation}
    The preparation covers the first two lines in \Cref{alg:discovery}.
    As previously discussed, the discovered \oclpmtext{}s should describe local patterns occurring in the process
    but also incorporate object interactions.
    Therefore, with the preprocessing, we extract dominant local dependencies between activities for each
    object type (place net discovery) on the one hand, and identify meaningful object interactions (event log
    transformation), on the other.

    \paragraph{Place Net Discovery}
    We model local dependencies between activities by using place nets, i.e., the simplest \lpmtext{}s containing
    only one place (see output of \emph{Preparation} in \Cref{fig:framework}).
    To obtain local dependencies for each object type, we flatten the event log for each object type (using $\flatocelfunc$
    from \Cref{subsec:preliminiaries-logs}) and execute a place net oracle $\placeoraclefunc$ on the flattened event log.
    As a place net oracle, we can use any discovery algorithm that returns a Petri net or a set of place nets.
    However, algorithms unrestricted to end-to-end trace fitness, as discussed in~\cite{DBLP:conf/apn/PeevaMA22}, are
    more suitable when we are interested in local dependencies.
    The result is the set $PT$, a set of place nets together with the object type for which they were discovered (Line 2
    in \Cref{alg:discovery}).

    \footnotetext{$\proj{\lpmvar}{p} = (\set{p}, T_p, F_p, l_p) \text{ such that } T_p = \set{t \in \gettuple{\lpmvar}{T}
    \st (t, p) \in \gettuple{\lpmvar}{F} \lor (p, t) \in \gettuple{\lpmvar}{F}},
    F_p = \set{(x, y) \in \gettuple{\lpmvar}{F} \st x = p \lor y = p} \text{, and } l_p = \proj{\gettuple{\lpmvar}{l}}{
        T_p}$}

    \paragraph{Event Log Transformation}
    To focus on object interactions, we enhance the event log with a new object type, used to group interacting objects,
    whose goal is to mimic the case notion from traditional event logs.
    Each object of the new object type represents a group of original objects that have met during the process.
    By then flattening the event log on the newly added object type, we extract process executions representing the process
    from the viewpoint of the interacting objects.
    In \Cref{alg:discovery}, we compute such object type and enhance the event log with it, by assuming a process
    execution oracle (see $\processexecutionfunc$ in Line 3) and extract the process executions by flattening
    (see $\flatocelfunc$ in Line 3).
    One way of discovering such object types and extracting process executions has been proposed
    in~\cite{DBLP:conf/icpm/AdamsSSSA22}.
    There are a multitude of ways one can define object interactions.
    One example is to consider sharing an event an interaction.
    Moreover, the new object type must not combine all interacting objects.
    It is up to the process execution oracle to decide what is an interaction and where to put the boundaries between the
    interacting objects.
    By abstracting from the concrete computation, we allow flexibility when it comes to which object interactions are
    meaningful.

    \subsection{Discovery (Phase 2)}\label{subsec:oclpm-discovery}
    In the discovery step, the event log given as input and the intermediate results from the preparation are used to
    build \oclpmtext{}s.
    First, we discover traditional \lpmtext{}s (\lpmtext{} discovery), and then construct the \oclpmtext{}s by enhancing the
    discovered \lpmtext{}s with place to object type mapping (object type annotation) and variable arcs (variable arc
    identification).
    Below, we describe each in more detail and finish the algorithm by returning the set of computed \oclpmtext{}s.

    \paragraph{LPM Discovery}
    We discover traditional \lpmtext{}s on the simple event log we created by utilizing an existing \lpmtext{} discovery
    technique~\cite{DBLP:conf/apn/PeevaMA22}.
    The technique used, starts with a precomputed set of local dependencies, represented as place nets (Line 5 in
    \Cref{alg:discovery}), and merges those into larger \lpmtext{}s only when there is evidence in the provided event
    log that the \lpmtext{} occurs in the process.
    Additionally, the occurrence of the \lpmtext{} should be for interacting objects.
    Let us consider the \oclpmtext{} in \Cref{fig:framework}.
    A claim that the \oclpmtext{} occurred in the event log means it occurred for related
    packages and items, and not random pairs of items and packages.
    The discovered set of \lpmtext{}s $\lpmsetvar$ (Line 6 in \Cref{alg:discovery}) satisfies this requirement because the
    simple event log focuses on interacting objects as explained above.

    \paragraph{Object Type Annotation}
    An important advantage of \oclpmtext{}s is depicting object type interactions.
    Therefore, for each place of an \lpmtext{}, we identify the object type they represent.
    Since we used the computed place nets for each object type $P$ as a starting point for the \lpmtext{} discovery, we
    use the original place net to object type mapping $PT$ to compute the place to type mapping (Line 9
    in \Cref{alg:discovery}).

    \paragraph{Variable Arc Identification}
    Variable arcs allow for modelling many-to-one interactions between objects of different types.
    Therefore, the final step is for each \lpmtext{} to identify the variable arcs.
    In our approach, we identify variable arcs as proposed in~\cite{DBLP:journals/fuin/AalstB20}.
    More concretely, given an \lpmtext{} (which is a labeled Petri net) $\lpmvar = (P, T, F, l)$, we define
    $\netvararcsvar = \set{(p, t)
        \in \netarcsvar \cap (P \times T) \st \vararcscorefunc(l(t), \enhancedplacenetsset(p)) < \tau} \cup \set{(t, p)
        \in \netarcsvar \cap (T \times P) \st \vararcscorefunc(l(t), \enhancedplacenetsset(p)) < \tau}$, where
    $\vararcscorefunc \in \actuniverse \times \otuniverse \partmapsto \multiset{0,1}$
    computes the fraction of events of the specified activity that contain exactly one object of the object type in
    question, as defined below, and $\tau$ is a user-defined threshold.

    \begin{equation*}
        \vararcscorefunc(\eactvar, \otvar) = \frac{\length{\set{\eventvar \in \eventsetvar_{\oclpmvar} \st
        \geteventattr{\eventvar}{act} = \eactvar \land \length{\geteventattr{\eventvar}{obj}(\otvar)} = 1}}}
        {\length{\set{\eventvar \in \eventsetvar_{\oclpmvar} \st \geteventattr{\eventvar}{act} = \eactvar}}}
    \end{equation*}

    Note, for the variable arc computation, we use $\eventsetvar_{\oclpmvar}$ since \oclpmtext{}s do not cover all
    events in an event log.
    Finally, the $\variablearcfunc$ in Line 10 returns $\netvararcsvar$ computed as described before.

    With this we covered all steps of the discovery algorithm.

    \section{Evaluation}\label{sec:evaluation}In this section, we evaluate the proposed approach qualitatively and quantitatively.
    First, we demonstrate applicability on two case studies.
    Then, we report runtime statistics and count of discovered models for different event logs.
    This is the first method for discovering \oclpmtext{}s, so we are unable to compare with previous work.
    All experiments are performed using the open-access implementation we provide in the ProM
    framework\footnote{\url{https://github.com/promworkbench/ObjectCentricLPMs}}.
    As a place net oracle, we use the SPECpp plugin in ProM\footnote{\url{https://github.com/promworkbench/SPECpp}}, and
    as an \lpmtext{} discovery approach we use the approach from~\cite{DBLP:conf/apn/PeevaMA22} also available in
    ProM\footnote{\url{https://github.com/promworkbench/LocalProcessModelDiscoveryByCombiningPlaces}}.
    For each event log, we build \oclpmtext{}s between $2$ and $7$ places, $3$ and $10$ transitions, all activities and
    discovered place nets, and window size $7$.

    \subsection{Case Studies}\label{subsec:eval-case-studies}

    \subsubsection{BPI Challenge 2017}

    \begin{figure}[t]
        \centering
        \includegraphics[width=0.8\linewidth]{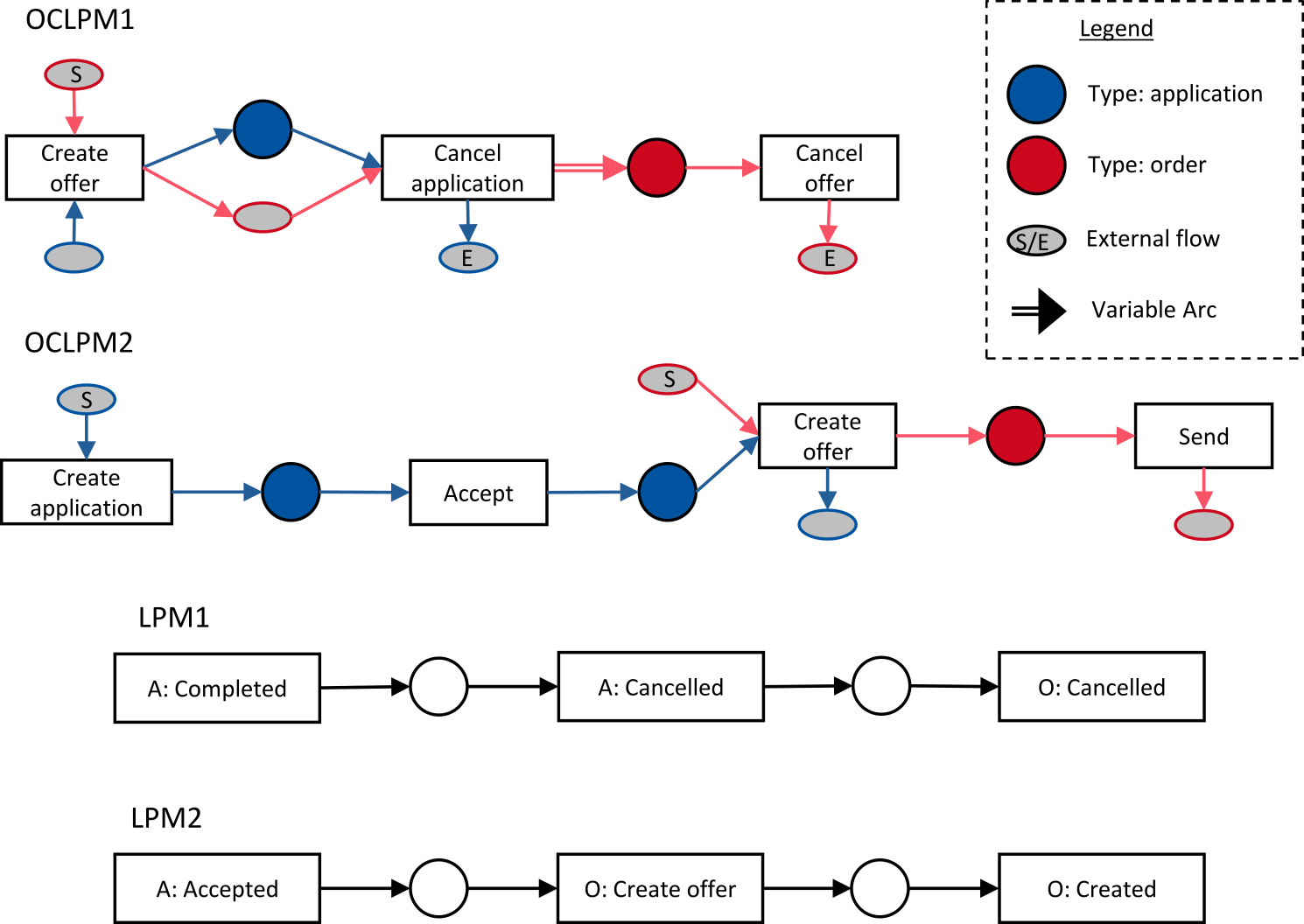}
        \caption{Discovered \oclpmtext{}s with the proposed approach and \lpmtext{}s with
        ~\cite{DBLP:conf/apn/PeevaMA22} on the BPIC2017 event log. In \oclpmtext{}s, we model external flow of the
        objects with elipses, where S/E denote object Start/End position.}
        \label{fig:eval_casestudy2}
    \end{figure}

    The event log is recorded from a loan application process of a Dutch financial institute~\cite{bpic2017}.
    It involves objects of types \emph{application} and \emph{offer}.
    The event log is available in both traditional xes and \oceltext{} format, hence we use it to compare \oclpmtext{}
    discovery to \lpmtext{} discovery on the same event log.

    We discover \lpmtext{}s using~\cite{DBLP:conf/apn/PeevaMA22} on the traditional event log, and \oclpmtext{}s on the
    available \oceltext{} with the proposed approach.
    In \Cref{fig:eval_casestudy2}, we display two \oclpmtext{}s and two \lpmtext{}s.
    The \oclpmtext{} at the top depicts how after an application is created and then accepted, a new offer is created and
    sent.
    The interaction between the application and offer object types is clearly shown in the discovered \oclpmtext{}.
    The highest-ranked \lpmtext{} also shows the move from application to offer, denoted with activity names prefixed
    with A and O.
    However, this is achieved with preprocessing of the event log.
    Emphasizing object types visually and treating them as first-class citizens gives a much clearer picture of what the
    pattern is describing.
    Moreover, it is not just the visual appeal that is gained with \oclpmtext{}s, but also expressiveness.
    This is illustrated with the help of the \oclpmtext{} and \lpmtext{} concerned with a cancellation of application.
    The shown \oclpmtext{}, with the help of variable arcs, clearly illustrates that one application might connect to
    multiple offers.
    Meaning, once the application is cancelled, all of the offers should be cancelled as well.
    The discovered \lpmtext{}, although depicting that offer should be cancelled after an application is cancelled, does
    not contain information on whether one or multiple offers are cancelled.

    \subsubsection{Order Management}

    \begin{figure}[t]
        \centering
        \begin{subfigure}[t]{0.35\textwidth}
            \centering
            \includegraphics[width=\linewidth]{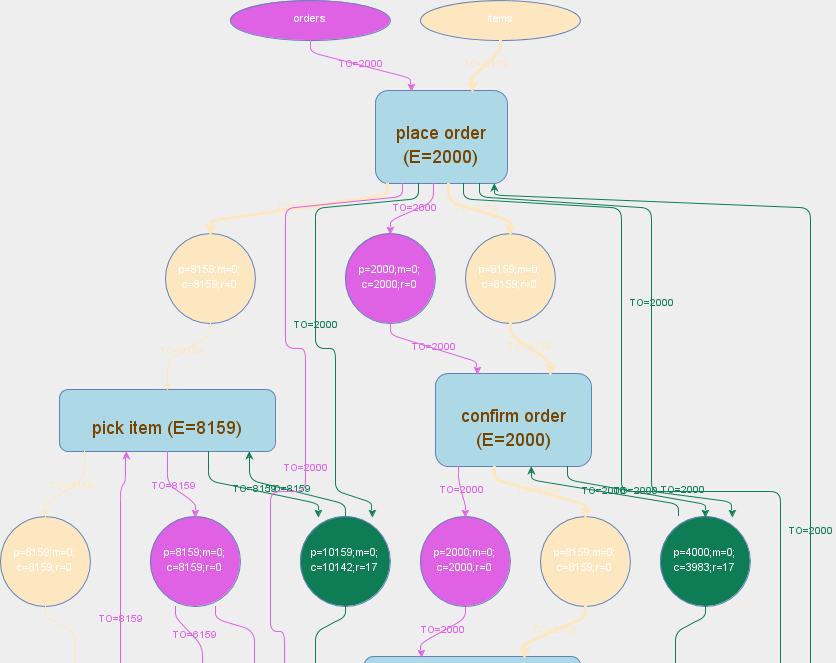}
            \caption{Part of the full end-to-end process model describing the same behavior as the \oclpmtext{}.}
            \label{fig:eval_casestudy1-ocpm}
        \end{subfigure}
        \hspace{0.05cm}
        \begin{subfigure}[t]{0.62\textwidth}
            \centering
            \includegraphics[width=\linewidth]{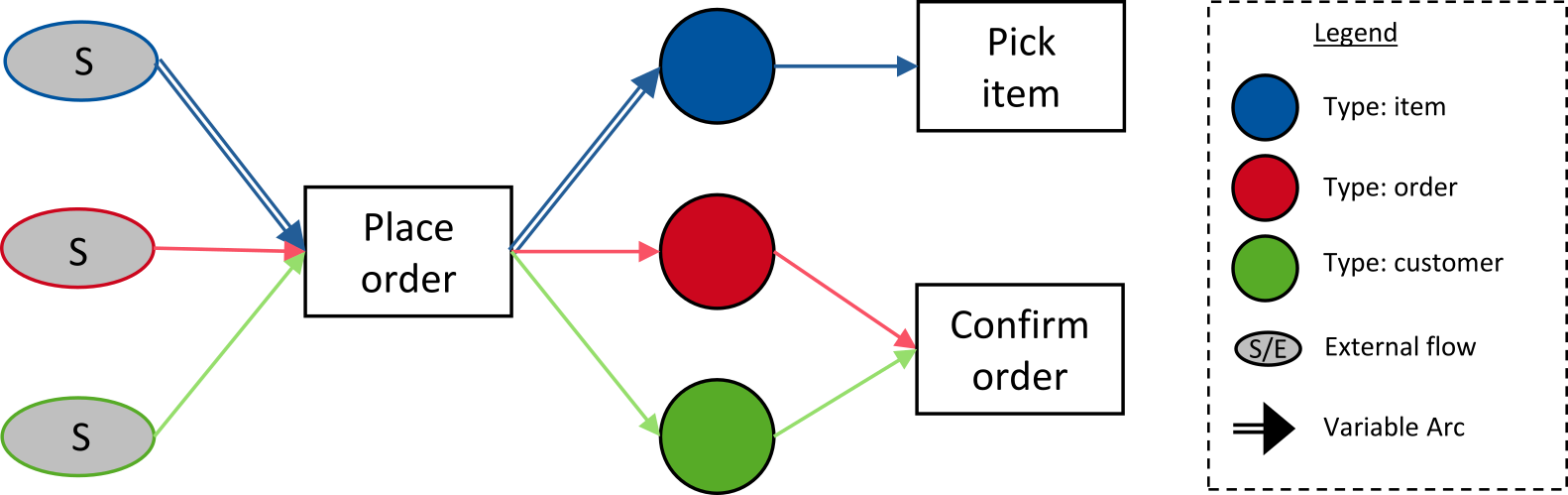}
            \caption{Discovered \oclpmtext{} showcasing the interactions between items, orders, and customers,
                focusing on the activities \emph{place order}, \emph{pick item}, and \emph{confirm order}.}
            \label{fig:eval_casestudy1-oclpm}
        \end{subfigure}
        \caption{Part of the end-to-end model and one \oclpmtext{} for the Order Management event log.}
        \label{fig:eval_casestudy1}
    \end{figure}

    The event log, as its name hints, describes a process for managing orders in which objects of types
    \emph{order}, \emph{item}, \emph{package}, \emph{customer}, and \emph{product} are involved.
    The main flow of the process is that a customer places an order, which is then confirmed by an employee, continuing
    into two disentangled subprocesses.
    On the one hand, we have, collecting order items, packing them, and sending the package (possibly multiple times if
    the deliveries have failed) until a successful delivery, and on the other hand, paying the order and possibly sending
    multiple reminders before the payment was completed.

    The approach proposed run about $40$ seconds and discovered in total $375$ models.
    We show the highest ranked \oclpmtext{} in \Cref{fig:eval_casestudy1-oclpm}.
    The model clearly shows the relationship between orders, customer, and items.
    One order is made by one customer, and one order can correspond to multiple items.
    Furthermore, it shows that \emph{place order} is a starting activity for orders, customers, and items.

    The end-to-end process model discovery took about $4$ seconds to discover the original model.
    Although, the process itself is not too complex, the model was very cluttered and spaghetti like.
    In \Cref{fig:eval_casestudy1-ocpm}, we show a part of the end-to-end process model.
    We had to filter on the orders, customers, and items object types, the activities, and discard most of the paths, to
    spot the relationships described by the \oclpmtext{} in \Cref{fig:eval_casestudy1-oclpm}, despite it being frequent and
    highly-ranked.
    The approach returned additional $374$ \oclpmtext{}, bringing valuable information that otherwise would have been
    lost in the complexity of the end-to-end model.

    \subsection{Results for Other Event Logs}\label{subsec:eval-statistics-on-different-event-logs}

    \begin{table}[t]
        \centering
        \caption{Event log description with number of discovered OCLPMs and runtime\protect\footnotemark.}
        \begin{tabulary}{\textwidth}{|C|C|C|C|C|C|}
            \hline
            Name & Events & Objects & Object Types & Models & Runtime(s)
            \\ \hline
            Order Management & 22367 & 11521 & 5 & 846 & 161 \\ \hline
            O2C & 98350 & 107767 & 19 & 1046 & 258 \\ \hline
            P2P & 24854 & 74489 & 8 & 2923 & 195 \\ \hline
            Transfer & 10319 & 2500 & 5 & 8 & 6 \\ \hline
            Recruiting & 6980 & 1505 & 6 & 109 & 43 \\ \hline
            Github & 1798 & 532 & 3 & 2314 & 92 \\ \hline
            BPIC2017 & 31203 & 8416 & 2 & 918 & 16 \\ \hline
        \end{tabulary}
        \label{tab:event-logs}
    \end{table}
    \footnotetext{{\url{https://www.ocel-standard.org/1.0/\#eventlogs}}}

    In this section, we report different statistics regarding the discovery of \oclpmtext{}s on various \oceltext{}.
    In \Cref{tab:event-logs} we list the event logs used in this part of the evaluation together with the number of
    events, objects, and object types they include.
    We run the \constname{Object-Centric Local Process Model Discovery given OCEL} plug-in in \constname{ProM} with
    default parameters.
    In \Cref{tab:event-logs}, we report the number of models discovered and the time necessary to do so for each of the
    event logs.
    The three largest event logs \constname{Order Management}, \constname{O2C}, and \constname{P2P} have the highest
    running time.
    We can also note that usually the runtime proportionally increases with the number of models built.
    One exception is the \constname{Github} event log, that has more \oclpmtext{}s discovered in less time, compared to
    the \constname{O2C} and the \constname{Order Management} event logs.
    However, \constname{Github} has significantly fewer events and objects than those two event logs.

    \section{Discussion}\label{sec:discussion}
    \begin{figure}[t]
        \centering
        \includegraphics[width=\linewidth]{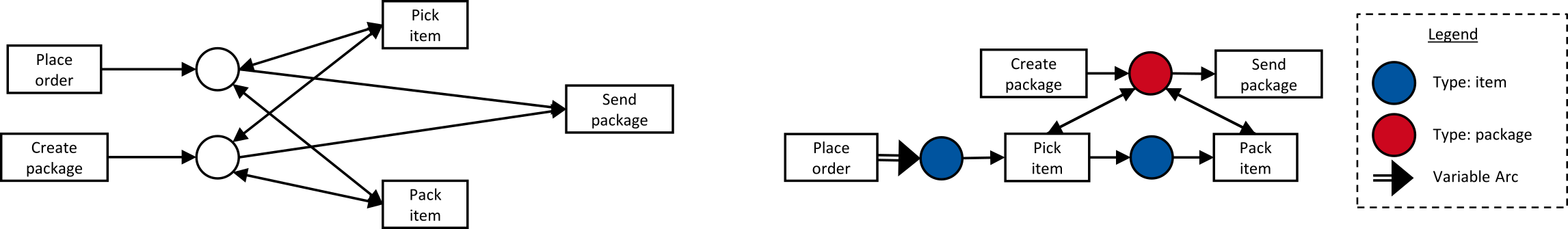}
        \caption{Example \lpmtext{} with missing (dashed line) and improper (red) dependencies, and \oclpmtext{} for
        the motivational example log.}
        \label{fig:lpm_vs_oclpm}
    \end{figure}

    In this section, we cover the strengths and weaknesses of the proposed approach.
    An alternative to our approach, would be to directly use traditional \lpmtext{} discovery on the flattened version of
    the event log and skip all the other steps we presented.
    In \Cref{fig:lpm_vs_oclpm} we show one \lpmtext{} and one \oclpmtext{} discovered for the event log given in
    \Cref{sec:motivating-example}.
    The obvious arguments in favour of \oclpmtext{}s, also exhibited in the evaluation of the approach, are the
    explicitness of object types and the expressiveness of the variable arcs.
    However, examining further, we compare both models from the perspectives of convergence and divergence.
    First, note that the missing dependency between \emph{Pick item} and \emph{Pack item} in the \lpmtext{} is present in
    the \oclpmtext{}.
    This is due to the \emph{Preparation} step of our approach, in which we discover local dependencies per object type.
    Second, the improper dependency between \emph{Place order} and \emph{Send package} is avoided in the \oclpmtext{},
    again as a result of the \emph{Preparation} step.
    Both of these examples demonstrate \emph{divergence} problems for traditional \lpmtext{}s discovered for \oceltext{}s
    and the ability of our approach to avoid them.
    Additionally, the improper dependency also creates the untrue impression that for each \emph{Place order} a \emph{
        Send package} is executed, leading to \emph{convergence} problems.
    In the \oclpmtext{}, the two activities are executed for different object types, allowing the \oclpmtext{} to be
    matched to one package and two orders as included in the log.
    In conclusion, the proposed approach resolves the convergence and divergence problems that would be introduced if a
    traditional \lpmtext{} discovery was used.

    \section{Conclusion}\label{sec:conclusion}
    In this work, we introduce \oclpmtext{}s as OCPNs, and present a discovery algorithm for building them from \oceltext{}s.
    We adapt and utilize existing work on computing local dependencies between activities and
    \lpmtext{} discovery to support the \oclpmtext{} discovery.
    Moreover, we implement the proposed algorithm in the open-source process mining tool ProM and evaluate the usefulness of
    the proposed approach on one real-world and one artificial event log.
    Additionally, we report runtime statistics and discovered models on multiple event logs.
    Finally, we have discussed strengths and limitations of the proposed approach.

    Next steps in this area would be to enhance the discovered \oclpmtext{}s with object and event attributes or allow
    for guided discovery similar to traditional \lpmtext{}s.
    Furthermore, more elaborate filtering techniques for discarding or giving less weight on specific object types would
    allow more advanced discovery.
    Finally, exploring alternative methods for discovering \oclpmtext{}s would be valuable.

%
%
%
    \bibliographystyle{splncs04}
    \bibliography{bibliography}

\end{document}